\documentclass[conference]{IEEEtran}
\usepackage[letterpaper, left=1.02in, right=1.02in, bottom=1in, top=0.75in]{geometry}
\usepackage{cite}
\usepackage{amsmath,amssymb,amsfonts}
\usepackage[pdftex]{graphicx}
\usepackage{color}
\usepackage{caption,subcaption}
\usepackage{array, booktabs, multirow}
\newcommand{\PreserveBackslash}[1]{\let\temp=\\#1\let\\=\temp}
\usepackage{algorithm}
\usepackage{algpseudocode}

\renewcommand{\algorithmicrequire}{\textbf{Input:}}
\renewcommand{\algorithmicensure}{\textbf{Output:}}

\definecolor{myred}{rgb}{1,0.2,0.2}
\definecolor{myblue}{rgb}{0,0.3,1}
\definecolor{mygreen}{rgb}{0.2,0.7,0}
\definecolor{myorange}{rgb}{1,0.5,0}

\usepackage{textcomp}
\def\BibTeX{{\rm B\kern-.05em{\sc i\kern-.025em b}\kern-.08em
    T\kern-.1667em\lower.7ex\hbox{E}\kern-.125emX}}
\begin{document}

\title{{Light-Weight DDoS Mitigation at Network Edge with Limited Resources}
}

\author{
\IEEEauthorblockN{1\textsuperscript{st} Ryo Yaegashi}
\IEEEauthorblockN{3\textsuperscript{rd} Yu Nakayama}
\IEEEauthorblockA{
\textit{Department of Computer and Information Sciences} \\
\textit{Tokyo University of Agriculture and Technology}\\
Tokyo, Japan \\
\{ryo.y, yu.nakayama\}@ieee.org}
\and
\IEEEauthorblockN{2\textsuperscript{nd} Daisuke Hisano}
\IEEEauthorblockA{
\textit{Graduate School of Engineering}\\
\textit{Osaka University}\\
Osaka, Japan\\
hisano@ieee.org}
}

\maketitle

\begin{abstract}
The Internet of Things (IoT) has been growing rapidly in recent years.
With the appearance of 5G, it is expected to become even more indispensable to people's lives.
In accordance with the increase of Distributed Denial-of-Service (DDoS) attacks from IoT devices, DDoS defense has become a hot research topic.
DDoS detection mechanisms executed on routers and SDN environments have been intensely studied.
However, these methods have the disadvantage of requiring the cost and performance of the devices.
In addition, there is no existing DDoS mitigation algorithm on the network edge that can be performed with the low-cost and low performance equipments.
Therefore, this paper proposes a light-weight DDoS mitigation scheme at the network edge using limited resources of inexpensive devices such as home gateways.
The goal of the proposed scheme is to simply detect and mitigate flooding attacks.
It utilizes unused queue resources to detect malicious flows by random shuffling of queue allocation and discard the packets of the detected flows.
The performance of the proposed scheme was confirmed via theoretical analysis and computer simulation.
The simulation results match the theoretical results and the proposed algorithm can efficiently detect malicious flows using limited resources.
\end{abstract}


%
%
\section{Introduction}
The Internet of Things (IoT) has been rapidly growing over the past decade to change our lives and society.
Various and enormous smart embedded devices such as sensors are deployed everywhere and communicate with edge and/or cloud servers via the Internet.
This paradigm will be dominant in the era of 5G and beyond to support global services in a wide variety of areas including industry, education, logistics, government and so on.

Over the past years, Distributed Denial-of-Service (DDoS) attacks have been increasing in both frequency and volume~\cite{wang2018delving}.
DDoS is a network attack for disrupting legitimate services by flooding traffic to targeted servers.
IoT devices are used by malicious attackers for a powerful amplifying platform of large-scale cyber attacks~\cite{perakovic2015analysis, hoque2015botnet}.
A recent prominent example of such attacks is the Mirai botnet, which was identified in 2016~\cite{kolias2017ddos}.
Such botnets spread among IoT devices including consumer household devices by means of brute force~\cite{lyu2017quantifying}.
IoT botnets have continuously emerged by malwares such as Mirai variants~\cite{de2017analysis}.
DDoS defense against attacks from IoT devices is a significant research topic.

Many DDoS defense strategies have been proposed and developed.
Among them, DDoS mitigation at the network edge is a promising solution for reducing the amount of malicious traffic in the Internet.
The typical parameters used for monitoring incoming traffic to detect and identify malicious traffic are source IP address, traffic increasing degree, and similarity among the traffic~\cite{zhang2015communication}.
Also, packet-level machine learning technique for anomaly detection was proposed to distinguish normal and attack traffic~\cite{doshi2018machine}.
An edge computing approach was proposed to gather information about incoming traffic and communicate the collected information with nearby detection services~\cite{bhardwaj2018towards}.
A major drawback of these learning-based DDoS detection mechanisms is the cost for data collection and learning.
The processing load is also a significant issue for resource-limited devices at the network edge such as home gateways and inexpensive layer-2 switches.

To address this problem, this paper proposes a light-weight DDoS mitigation mechanism at the network edge using limited resources of inexpensive devices.
The goal of the proposed scheme is to simply detect and mitigate flooding attacks such as UDP flood.
It detects malicious flows by random shuffling of queue allocation, and the packets of detected flows are discarded without queuing in the same way as typical active queue management schemes.
It enables early mitigation of DDoS attacks at the network edge to prevent them from flowing into the network.
The proposed shuffling algorithm is simple and easy to execute in low-cost equipment without any additional hardware, because it leverages queue resources which are generally not fully utilized.
The contribution of this work is to propose a simple and light-weight DDoS mitigation mechanism which can be employed together with existing DDoS defense schemes.

This paper is organized as follows.
Related work and the contribution of this paper are introduced in section \ref{sec:rltd}.
The proposed DDoS mitigation scheme is explained in section \ref{sec:prp}.
The performance of the proposed scheme is theoretically analyzed in section \ref{sec:theo}.
The feasibility of the proposed idea is evaluated through computer simulations in section \ref{sec:sim}.
The conclusion of this paper is provided in section \ref{sec:cncl}.

%
%
\section{Related work} \label{sec:rltd}
There have been many research efforts on DDoS detection and mitigation.
A router-based defense mechanism called Pushback was developed in ~\cite{ioannidis2002implementing}, where functionality is added to each router to detect attack from congestion signature.
Perimeter-based defense mechanisms proposed in \cite{chen2005perimeter} enabled Internet service providers (ISPs) to provide anti-DDoS services to their customers.
In these mechanisms, edge routers cooperatively identify flooding sources to establish rate-limit filters to block attack traffic.
A collaborative defense mechanism was also proposed in \cite{lee2013codef} to enable routers to distinguish attack flows from legitimate flows.

DDoS mitigation has been a significant issue for software defined network (SDN) because of the centralized control architecture~\cite{kalkan2017defense}.
A distributed collaborative framework was presented in \cite{sahay2015towards} to allow customers to request DDoS mitigation service from ISPs.
In this framework, ISPs change the label of the anomalous traffic upon request and redirect them to security middleboxes, while attack detection and analysis modules are deployed at customer side.
In addition, a detection method by using the entropy change of the destination IP address was also proposed in \cite{mousavi2015early}.
For IoT environments, an enhanced distributed low-rate attack mitigating (eDLAM) mechanism was proposed~\cite{liu2019efficient} based on a game model to analyze the attack benefit between attacker and defender.

However, there have been few research on simple DDoS mitigation mechanisms that can easily be executed in low-cost devices at the network edge.
Complicated operations including pattern-matching and cooperative approaches are unsuitable for such resource-limited devices.
Therefore, the contribution of this paper is to propose a light-weight DDoS mitigation mechanism for the IoT era.
Note that the proposed scheme can be employed together with existing DDoS defense mechanisms mentioned above.
The detail of the proposed idea is introduced in the next section.

%
%
\section{Proposed DDoS mitigation} \label{sec:prp}
\subsection{Concept} \label{sec:prp_cncpt}
The goal of the proposed light-weight DDoS mitigation is to simply detect and mitigate flooding attacks such as UDP flood and HTTP flood at the network edge.
The main idea is utilizing queuing functions to enable DDoS mitigation using limited resources of low-cost nodes such as switches and gateways.
The nodes at the network edge detect malicious flows by random shuffling of queue allocation, and the packets of detected flows are discarded without queuing.

\begin{figure*}[!t]
\centering
  \subcaptionbox{Flow allocation \label{fig:seq1}}{\includegraphics[width=1.5in]{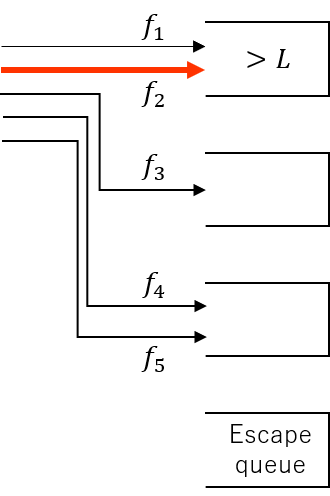}}
  \hspace{2.00em}
  \subcaptionbox{Reallocation \label{fig:seq2}}{\includegraphics[width=1.5in]{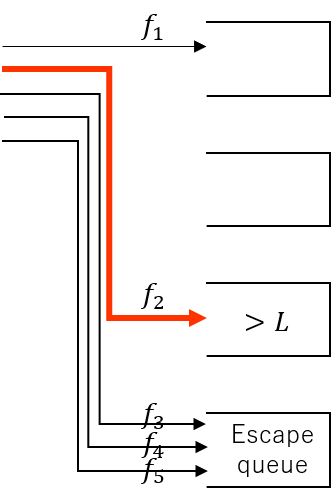}}
  \hspace{2.00em}
  \subcaptionbox{DDoS mitigation \label{fig:seq3}}{\includegraphics[width=1.5in]{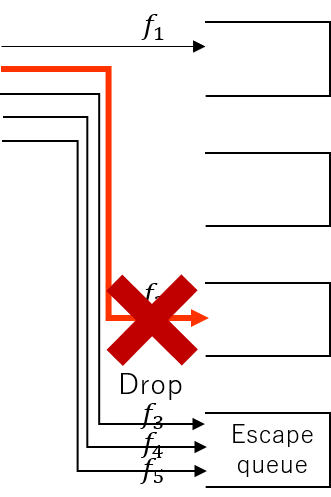}}
\caption{Concept of proposed DDoS mitigation.}
\label{fig:seq}
\end{figure*}

The proposed concept is depicted in Fig.~\ref{fig:seq}.
It is assumed that a node has several available queues for mitigation sequence.
An escape queue is configured for flow evacuation.
First, incoming flows are assigned to available queues as shown in Fig.~\ref{fig:seq1}.
If the enqueue rate for a queue exceeds a certain threshold ($L$), the flows ($f_{1}$ and $f_{2}$) are randomly reallocated to available queues.
Otherwise, allocated flows ($f_{3}$, $f_{4}$, and $f_{5}$) are detected as legitimate flows to be assigned to the escape queue as shown in Fig.~\ref{fig:seq2}.
After the reallocation process, the malicious flow can be identified and discarded to mitigate DDoS attacks, as illustrated in Fig.~\ref{fig:seq3}.

\subsection{Node architecture} \label{sec:prp_arch}
\begin{figure}[!t]
\centering
	\includegraphics[width=2.5in]{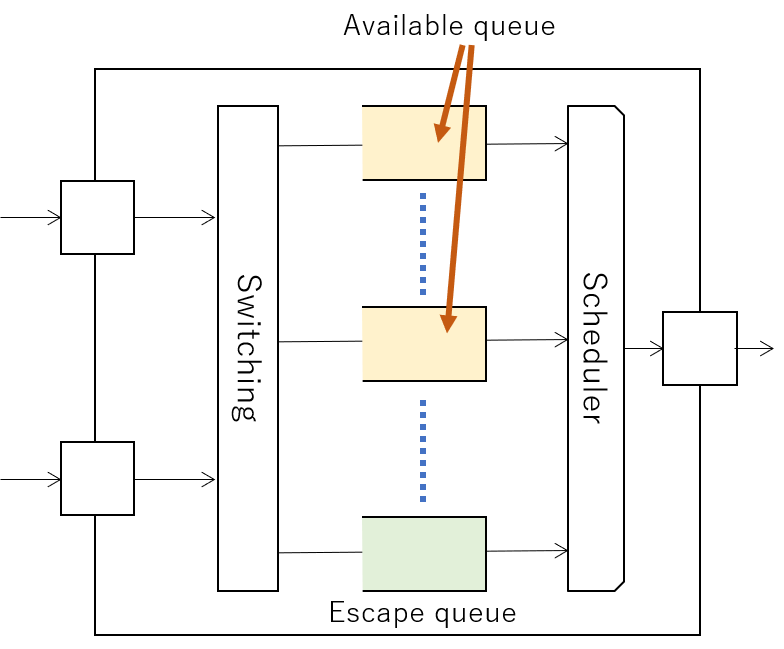}
	\caption{Node architecture.}
	\label{fig:arch}
\end{figure}

The assumed node architecture is depicted in Fig.~\ref{fig:arch}.
The requirement is to provide several queues, a switching function, and a scheduler, which is a general architecture for network nodes such as Ethernet switches.
For instance, many Ethernet switches provide eight queues for ensuring QoS among priority classes of traffic, whereas many of them are unused because generally two or three priority classes are used to identify real-time and best effort traffic.
Therefore, a queue is configured as an escape queue from the unused queues.
The other queues are defined as available queues for the proposed mitigation process.

\subsection{Mitigation algorithm} \label{sec:prp_alg}
\subsubsection{Variable definition} \label{sec:prp_alg_var}
The variables used in the proposed algorithm are summarized in Table \ref{tbl:variables}.
The escape queue is described as $Q_{escape}$.
Let $\mathcal{Q}$ denote the set of available queues, and $q \in \mathcal{Q}$ is the queue identifier.
The current queue size is described as $l_{q}$.
The number of assigned flows to $q$th queue is $n_{q}$.
Also, $y_{q}$ is a boolean flag for $q$th queue.
Let $\mathcal{F}$ denote the set of incoming flows, and $f \in \mathcal{F}$ is the flow identifier.
The queue to which $f$th flow is assigned is denoted as $x_{f}$.
The threshold for increase in the queue size is defined as $L$.
Finally, $\tau$ is the cycle length.

\let\PBS=\PreserveBackslash
\begin{table}[!t]
	\renewcommand{\arraystretch}{1.2}
	\caption{Variables}
	\label{tbl:variables}
	\centering
	\begin{tabular}[t]{>{\PBS\centering\hspace{0pt}}p{0.5in} >{\PBS\centering\hspace{0pt}}p{2.0in}} \toprule
		Variable & Definition \\ \hline
		$Q_{escape}$ & Escape queue \\
		$\mathcal{Q}$ & Set of available queues \\
		$q$ & Queue identifier ($q \in \mathcal{Q}$) \\
		$l_{q}$ & Size of $q$th queue \\
		$n_{q}$ & Number of flows assigned to $q$th queue \\
		$y_{q}$ & Boolean flag for $q$th queue \\

		$\mathcal{F}$ & Set of incoming flows \\
		$f$ & Flow identifier ($f \in \mathcal{F}$) \\
		$x_{f}$ & Assigned queue for $f$th flow \\

		$L$ & Threshold for queue size \\
		$\tau$ & Cycle length \\

		\bottomrule
	\end{tabular}
\end{table}
\renewcommand{\arraystretch}{1}

\subsubsection{Algorithm} \label{sec:prp_alg_alg}
The proposed DDoS mitigation algorithm is described in Algorithm~\ref{alg:mitigation}.
It is cyclically executed with the time interval $\tau$.
\begin{algorithm}
\caption{DDoS mitigation algorithm}
\label{alg:mitigation}
\begin{algorithmic}[1]
	\State{\# Step 1: Queue size evaluation}
	\For{$q \in \mathcal{Q}$}
		\If{$l_{q} - l_{q}^{last} > L$}
			\State{$y_{q} \gets TRUE$}
		\Else
			\State{$y_{q} \gets FALSE$}
		\EndIf
		\State{$l_{q}^{last} \gets l_{q}$}
	\EndFor

	\State{\# Step 2: Queue reallocation}
	\For{$f \in \mathcal{F}$}
		\If{$y_{x_{f}} = TRUE$}
			\If{$n_{x_{f}} = 1$}
				\State{$x_{f} \gets Drop$}
			\Else
				\State{$x_{f} \gets Random(q \in \mathcal{Q})$}
			\EndIf
		\Else
			\State{$x_{f} \gets Q_{escape}$}
		\EndIf
	\EndFor
\end{algorithmic}
\end{algorithm}
The proposed algorithm consists of two steps.

In the first step, the increase in the queue size is evaluated for each available queue.
If the increase ($l_{q} - l_{q}^{last}$) exceeds the threshold $L$, the flag $y_{q}$ is set as \textit{TRUE}.
Otherwise, the flag is set as \textit{FALSE}.

In the second step, incoming flows are reallocated to available queues.
If the flag $y_{q}$ for the queue to which $f$th flow is assigned ($x_{f}$) is \textit{TRUE}, the flow is reallocated.
If $n_{x_{f}} = 1$, $f$th flow is detected as a malicious flow, and the node starts to drop it.
Otherwise, this flow is randomly assigned to available queues $q \in \mathcal{Q}$.
If $y_{x_{f}}$ is \textit{FALSE}, $f$th flow is detected as a legitimate flow.
Then, it is assigned to the escape queue $Q_{escape}$.

With this algorithm legitimate flows are allocated to the escape queue, while malicious flows are detected and dropped to mitigate DDoS attacks.

%
%
\section{Theoretical analysis} \label{sec:theo}
The performance of the proposed scheme is theoretically analyzed in this section.

\subsection{Accuracy for DDoS detection} \label{sec:theo_acc}
With the proposed algorithm, a malicious flow can be identified if only that flow is enqueued into an available queue.
Thus, the DDoS detection accuracy in a cycle is equal to the probability that only a malicious flow is assigned to an available queue.
Let $c$ denote the cycle identifier for executing Algorithm~\ref{alg:mitigation}.

Here we define $p_{q, c}$ as the probability that only a flow is assigned to $q$th queue at $c$th cycle.
Since the allocation probability for each queue is uniform, $p_{q, c}$ is described as
\begin{equation}
p_{q, c} = \frac{1}{N_{Q}} \left( \frac{N_{Q} - 1}{N_{Q}} \right)^{N_{F, c} - 1},
\label{eq:pqc}
\end{equation}
where $N_{Q} = | \mathcal{Q} |$ is the number of available queues and $N_{F, c}$ denotes the number of flows that satisfy $x_{f} = q \in Q$, i.e. non-escaped flows, at $c$th cycle.

Let $p_{c}$ denote the probability that a flow is assigned to any of available queues and no other flow is assigned to the same queue at $c$th cycle, which is calculated as
\begin{equation}
\begin{split}
p_{c} &= \sum_{q \in \mathcal{Q}} p_{q, c} \\
      &= \left( \frac{N_{Q} - 1}{N_{Q}} \right)^{N_{F, c} - 1}.
\label{eq:pc}
\end{split}
\end{equation}

Let $U_{c}$ and $E_{c}$ denote the expected number of undetected and detected malicious flows at $c$th cycle, respectively.
The relationship between these variables is 
\begin{equation}
U_{c+1} = Max(0, U_{c} - E_{c}).
\label{eq:uc+1}
\end{equation}
Note that $U_{0} = \rho N_{F,0}$ where $\rho$ denotes the ratio of malicious flows to all incoming flows.
$E_{c}$ is described as
\begin{equation}
E_{c} = U_{c} p_{c}.
\label{eq:ec}
\end{equation}

\subsection{Escape probability} \label{sec:theo_esc}
Assuming that the threshold $L$ is sufficiently large, all the flows that are not enqueued to the same queue with malicious flows can be assigned to the escape queue.
The increase in the number of escaped flows improves the DDoS detection accuracy as formulated in \eqref{eq:ec}.

The expected number of assigned flows to a queue at $c$th cycle is simply calculated as
\begin{equation}
n^{ex}_{q, c} = \frac{N_{F, c}}{N_{Q}}.
\label{eq:nex}
\end{equation}
The probability that there is no malicious flow among them is
\begin{equation}
\gamma_{q, c} = \left( 1 - \frac{U_{c}}{N_{F, c}} \right)^{n^{ex}_{q, c}}.
\label{eq:gamma}
\end{equation}

Therefore, the expected value of escaped flows at $c$th cycle is
\begin{equation}
\begin{split}
V_{c} &= \sum_{q \in \mathcal{Q}} n^{ex}_{q, c} \gamma_{q, c} \\
      &= N_{F, c} \gamma_{q, c}.
\label{eq:vc}
\end{split}
\end{equation}

\subsection{Time variation} \label{sec:theo_time}
From the number of non-escaped flows $N_{F, c}$, the numbers of detected malicious flows and newly escaped flows are reduced at the next cycle.
Thus, $N_{F, c+1}$ is described as
\begin{equation}
N_{F, c+1} = Max(0, N_{F, c} - E_{c} - V_{c}).
\label{eq:nfc+1}
\end{equation}
Since both $E_{c}$ and $V_{c}$ are the functions of $N_{F, c}$, the DDoS detection accuracy at each cycle can be computed with these equations.

%
%
\section{Computer simulation} \label{sec:sim}
\subsection{Simulation condition} \label{sec:sim_cnd}
The performance of the proposed scheme was confirmed with computer simulation.
First, we verified the consistency with the theorical performance formulated in \eqref{eq:ec} and \eqref{eq:nfc+1}.
The number of incoming flows $|\mathcal{F}|$ was set to $16$, $32$, and $64$ to evaluate the detection accuracy in different conditions.
The number of available queues was set as $N_{Q} = 6$.

Second, the performance in dynamic conditions was evaluated.
The number of available queues was set as $N_{Q} = 7$.
The number of incoming flows $|\mathcal{F}|$ was $64$ in the initial condition, and additional flows arrive during $1 \leq c \leq 15$.
The number of additional flows was set as $12$ and $24$ per cycle.
All adding flows were assumed to be legitimate flows.
From \eqref{eq:vc}, the expected value of escaped flows is calculated as $V_{c} = 24$ when $N_{Q} = 7$ and $|\mathcal{F}| = 64$; the performance limitation of the proposed scheme for this configuration is $24$ additional flows per cycle.

Finally, to clarify the marginal performance, the relationship between the numbers of incoming flows and available queues was evaluated.
We define the processed flows as the total of escaped flows and detected malicious flows.
We set the number of incoming flows $|\mathcal{F}| = 128$.
The number of processed flows in $30$ cycles was measured with different number of available queues.

The other parameters commonly used in these simulations were as follows.
Threshold for queue size $L$ was set at $62.5$ MB.
The upstream link bandwidth was $50$ $\mathrm{Mbps}$.
The enqueued data size of legitimate flows during each cycle was randomly determined using a normal distribution with a mean $2.0$ Mbps and a standard deviation $0.5$ Mbps.
The data rate of malicious flows was set as $100$ Mbps assuming UDP flooding attacks.
The ratio of malicious flows in all incoming flows was set at $\rho = 0.1$.
The cycle length $\tau$ was $10$s, and the total simulation time was $300$s.
The simulation was iterated $1000$ times for each condition.
The parameters explained above are summarized in Table \ref{tbl:para}.

\let\PBS=\PreserveBackslash
\begin{table}[!t]
	\renewcommand{\arraystretch}{1.2}
	\caption{Simulation parameters}
	\label{tbl:para}
	\centering
	\begin{tabular}[t]{>{\PBS\centering\hspace{0pt}}p{1.25in} >{\PBS\centering\hspace{0pt}}p{1.25in}} \toprule
		Parameter & Value \\ \hline
    Threshold $L$ & $62.5$ $\mathrm{MB}$ \\
    Link speed & $50$ $\mathrm{Mbps}$ \\
    Ratio of DDoS $\rho$ & $0.1$\\
    Cycle length $\tau$ & $10$ s \\
    Simulation time & $300$ s \\
    Number of iterations & $1000$ \\
		\bottomrule
	\end{tabular}
\end{table}
\renewcommand{\arraystretch}{1}

\begin{figure*}[!t]
\centering
  \subcaptionbox{Incoming flows, $|\mathcal{F}| = 16$ \label{fig:DDoS_F16}}{\includegraphics[width=2.0in]{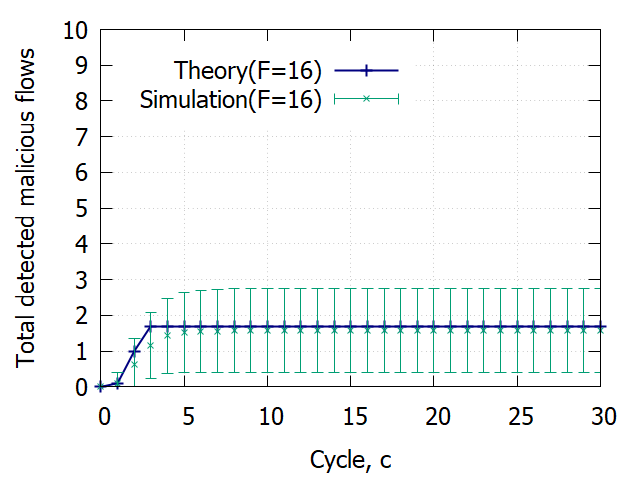}}
  \hspace{1.00em}
  \subcaptionbox{Incoming flows, $|\mathcal{F}| = 32$ \label{fig:DDoS_F32}}{\includegraphics[width=2.0in]{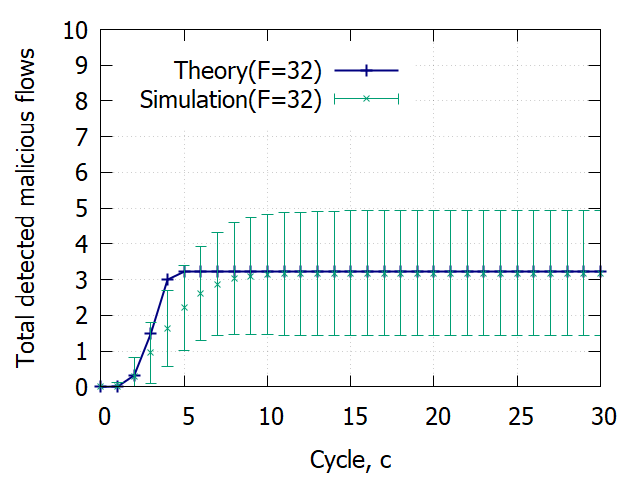}}
  \hspace{1.00em}
  \subcaptionbox{Incoming flows, $|\mathcal{F}| = 64$ \label{fig:DDoS_F64}}{\includegraphics[width=2.0in]{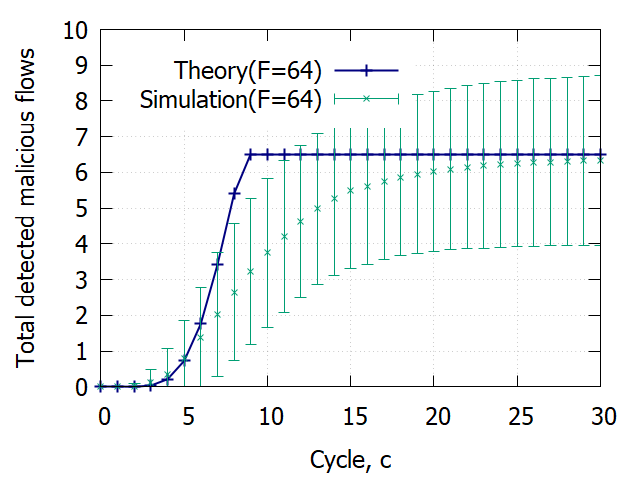}}
\caption{Number of detected DDoS flows, $E_{c}$}
\label{fig:detected DDoS}
\end{figure*}

\begin{figure*}[!t]
\centering
  \subcaptionbox{Incoming flows, $|\mathcal{F}| = 16$ \label{fig:N_F,c_F16}}{\includegraphics[width=2.0in]{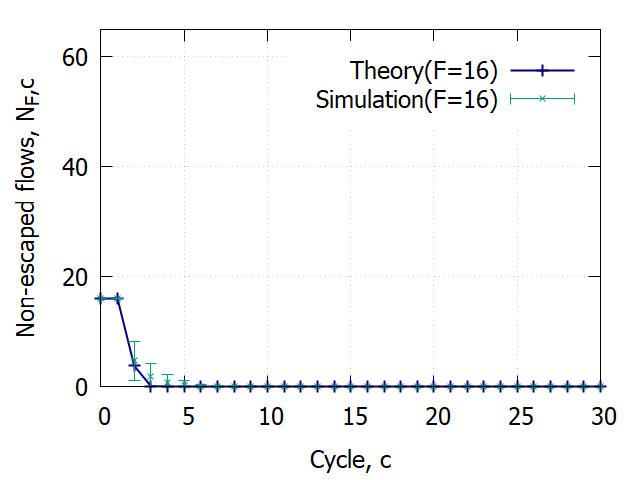}}
  \hspace{1.0em}
  \subcaptionbox{Incoming flows, $|\mathcal{F}| = 32$ \label{fig:N_F,c_F32}}{\includegraphics[width=2.0in]{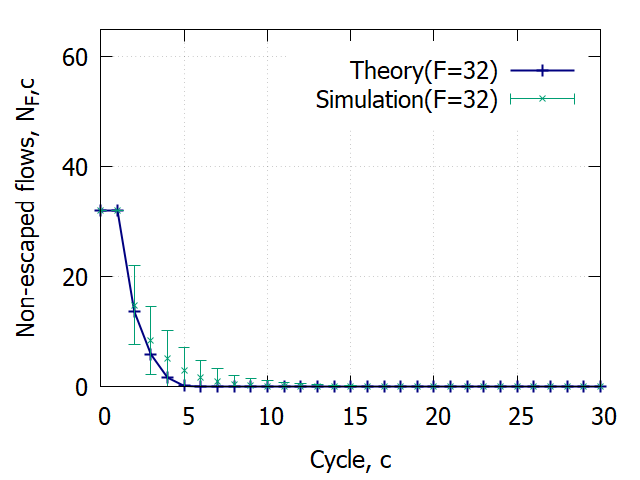}}
  \hspace{1.0em}
  \subcaptionbox{Incoming flows, $|\mathcal{F}| = 64$ \label{fig:N_F,c_F64}}{\includegraphics[width=2.0in]{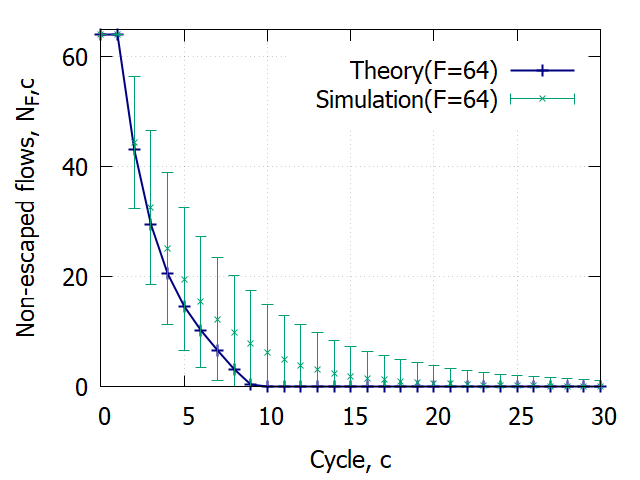}}
\caption{Number of non-escaped flows, $N_{F, c}$}
\label{fig:N_F,c}
\end{figure*}

\subsection{Simulation results} \label{sec:sim_rslt}
Figs.~\ref{fig:DDoS_F16}--\ref{fig:DDoS_F64} show the theoretical and simulated values of detected malicious flows in each cycle with $|\mathcal{F}|$ = $16$, $32$, and $64$.
Figs.~\ref{fig:N_F,c_F16}--\ref{fig:N_F,c_F64} show the theoretical and simulated values of non-escaped flows in each cycle.
The points represent the average number of detected malicious flows and non-escaped flows, and the errorbars represent the standard deviation.
We can see that the theoretical values are settled within the ranges at any cycle.
Note that the variation between the theoretical and simulated results is due to the randomness of the simulation.
Thus, it was confirmed from these results that the proposed scheme properly works to detect and mitigate DDoS attacks.

\begin{figure*}[!t]
\centering
  \subcaptionbox{$12$ additional flows per cycle \label{fig:add12}}{\includegraphics[width=3.0in]{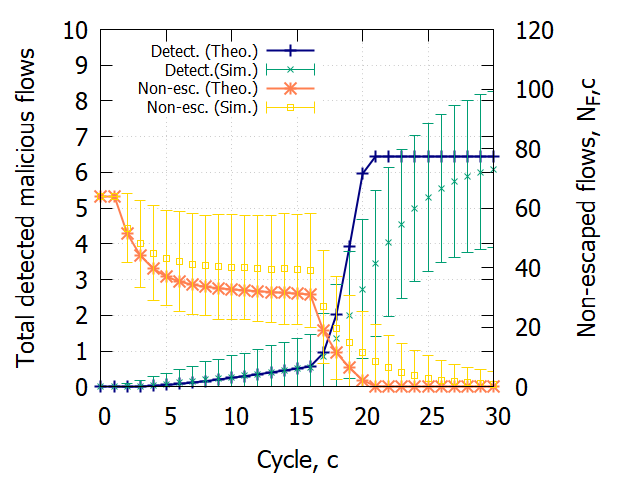}}
  \hspace{1.0em}
  \subcaptionbox{$24$ additional flows per cycle \label{fig:add24}}{\includegraphics[width=3.0in]{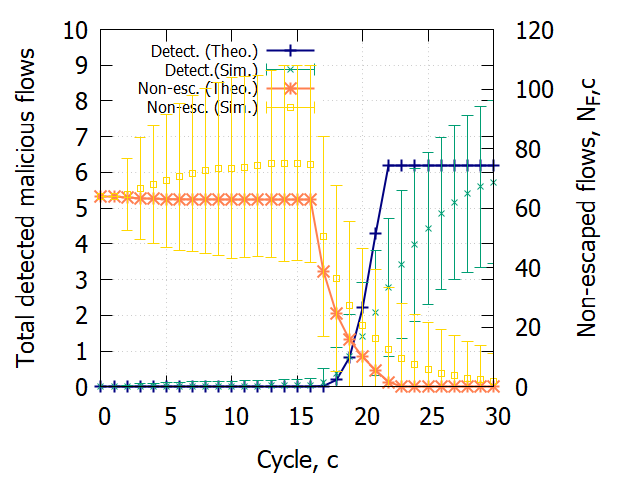}}
\caption{Simulation results in dynamic conditions}
\label{fig:dynamic}
\end{figure*}

Figs.~\ref{fig:add12}--\ref{fig:add24} show the theoretical and simulated results for dynamic conditions.
The points represent the average number of detected malicious flows and non-escaped flows, and the errorbars represent the standard deviation.
The simulation results match the theoretical values.
When $12$ flows were added per cycle, they can be gradually escaped with the proposed algorithm.
Since the performance limit with $Q_{N} = 7$ is $24$ escaped flows per cycle, the proposed scheme holds on during $c \leq 15$ and DDoS detection proceeds after $15$th cycle as shown in Fig.~\ref{fig:add24}.
These results imply that the proposed scheme works even in dynamic environments if the number of additional flows is less than the performance limit.

\begin{figure}[!t]
\centering
	\includegraphics[width=2.5in]{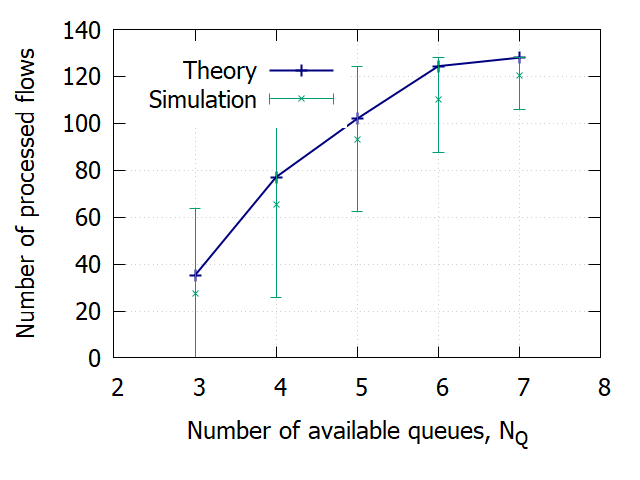}
	\caption{Number of processed flows}
	\label{fig:limit}
\end{figure}

Fig.~\ref{fig:limit} shows the theoretical and simulated values of processed flows after the $30$th cycle with $N_{Q}$ = $3$, $4$, $5$, $6$ and $7$.
The number of processed flows increases in accordance with the increase of available queues.
Most of the flows are escaped or dropped when $N_{Q} = 7$.
We can also see that about $85$\% of the incoming flows are assigned to the escape queue even when the number of available queues is $6$.
Hence, it is thought that more than $100$ flows can be identified in $30$ cycles if the number of available queues is greater than $5$.
It is implied that the proposed scheme provides enough scalability for the network edge focused in this paper.

%
%
\section{Conclusion} \label{sec:cncl}
In recent years, the number of DDoS attacks has been increasing.
DDoS defense mechanisms against attacks from IoT devices have become an important issue, and various methods have been proposed and developed.
However, there have been no DDoS mitigation schemes for low-cost and low-performance devices at the network edge such as home gateways.
To address this problem, in this paper we proposed a lightweight DDoS mitigation mechanism at the network edge using the limited resources of inexpensive devices.
This mechanism detects malicious flows by random shuffling of queue allocation, and the packets of detected flows are discarded without queuing.
It enables simple detection and mitigation of flooding attacks such as UDP flood.
It was confirmed through computer simulation that the proposed algorithm identified more than $100$ flows by using $7$ queues for queue assignment.
With the proposed scheme, DDoS attacks can be simply mitigated even in low-cost devices with limited resources.

%
%
\section*{Acknowledgment}
A part of this work This work was supported by JST, ACT-I, Grant Number JPMJPR18UL, Japan.

%
%
\bibliographystyle{IEEEtran.bst}
\bibliography{bibliography}

\end{document}